\documentclass[conference]{IEEEconf}

\input epsf
\usepackage{graphicx}
\usepackage{multirow}
\usepackage{titlesec}
\usepackage{balance} 
\usepackage{stfloats}
\usepackage{xcolor}
\usepackage{cite}
\usepackage{amsmath,amssymb,amsfonts}
\usepackage{algorithmic}
\usepackage{graphicx}
\usepackage{textcomp}
\usepackage{xcolor}
\usepackage{amsmath,amssymb,amsfonts}
\usepackage{algorithmic}
\usepackage{graphicx}
\usepackage{textcomp}
\usepackage{subcaption}
\usepackage{hyperref}
\usepackage{bm}
\usepackage{amsmath}
\usepackage{multirow}
\usepackage{booktabs}
\usepackage{multirow}
\usepackage{booktabs}
\usepackage{colortbl}
\usepackage{xcolor}
\usepackage{caption}
\usepackage{subcaption}
\usepackage{float} 
\usepackage{changepage} 
\usepackage[table]{xcolor}
\usepackage{booktabs}  % 三线表核心宏包
\usepackage{array}
\usepackage{tabularx}
\usepackage{tikz}
\usepackage{pgfplots}
\usepackage{fontawesome}
\usepackage{rotating}
\usepackage{pdflscape}
\pgfplotsset{compat=1.18}
\usepgfplotslibrary{statistics}
\usepackage{comment}
\usepackage{makecell}
\usepackage{fontawesome}
\usepackage{array}

\renewcommand\thesection{\arabic{section}} % arabic numerals for the sections

\renewcommand\thesubsectiondis{\thesection.\arabic{subsection}}% arabic numerals for the subsections

\renewcommand\thesubsubsectiondis{\thesubsectiondis.\arabic{subsubsection}}% arabic numerals for the subsubsections

\begin{document}

\title{\textbf{\Large Understanding How Enterprises Adopt the Model Context Protocol for LLM-Driven Software Engineering\\}}

\author{Kehui Chen$^{1}$, Yicheng Sun$^{1}$, Jacky Keung$^{1}$, Zhenyu Mao$^{1}$, Xiaoxue Ma$^{2,*}$\\
	\normalsize $^{1}$City University of Hong Kong, Hong Kong, China \\
	\normalsize $^{2}$Hong Kong Metropolitan University, Hong Kong, China \\
	\normalsize \{kehuichen2-c, yicsun2-c, zhenyumao2-c@my.cityu.edu.hk, Jacky.Keung@cityu.edu.hk, kxma@hkmu.edu.hk \\
	\normalsize *corresponding author
}

% \author{First A. Author$^{1}$, Second B. Author$^{2,*}$, and Third C. Author$^{3}$\\
% 	\normalsize $^{1}$First Affiliation, City, State, Country\\
% 	\normalsize $^{2}$Second Affiliation, City, State, Country\\
% 	\normalsize $^{3}$Third Affiliation, City, State, Country\\
% 	\normalsize author1@email.com, author2@email.com, author3@email.com\\
% 	\normalsize *corresponding author
% }
%+++++++++++++++++++++++++++++++++++++++++++

%\specialpapernotice{(Invited Paper)}

\maketitle
\begin{abstract}
Large Language Models (LLMs) are increasingly used in AI-based software engineering, but their limitations in complex task execution and multi-tool coordination have driven growing interest in the Model Context Protocol (MCP). Existing research has mainly focused on MCP's technical design, with limited empirical evidence on how it is adopted and used in enterprise practice, particularly with regard to deployment challenges, operational risks, and practitioner expectations. To address this gap, we conducted semi-structured interviews with 20 practitioners from eight companies in the Internet and financial sectors. The findings show that MCP is valued for supporting cross-system collaboration, task decoupling, and knowledge reuse in LLM-based workflows, but its adoption remains constrained by ecosystem fragmentation, cross-component coordination difficulties, and unresolved problems in distributed state management and fault diagnosis. Participants also expressed strong demand for better standardization, lower adoption barriers through low-code or plugin-based approaches, and more systematic operational support. These results provide early empirical evidence on enterprise MCP practice and offer practical implications for improving MCP's standardization, usability, and deployment readiness in real-world software engineering environments.
\end{abstract}
\IEEEoverridecommandlockouts
\vspace{1.5ex}

% for peerreview papers, inserts a page break and creates the second title.
% Will be ignored for other modes.
\IEEEpeerreviewmaketitle

\section{Introduction}

Large language models (LLMs) have been widely adopted in enterprise settings, but they still face notable limitations in handling complex tasks such as cross-system invocation and multi-tool collaboration \cite{hou2025model}. To address these challenges, Anthropic proposed the Model Context Protocol (MCP) in late 2024 \cite{anthropic_mcp_2024}. Since then, MCP has emerged as an intermediate-layer protocol for connecting LLMs with external tools and resources \cite{hou2025model}.  MCP is built around four core components—Host, Client, Server, and Resource—which together support standardized tool invocation and context exchange between LLMs and heterogeneous systems \cite{yang2025survey}.

Current research on MCP has primarily focused on three areas. First, benchmarking studies have proposed evaluation frameworks that combine MCP servers, evaluators, and extensible infrastructures to assess LLM performance on realistic tasks \cite{wang2025mcp}. Second, performance-oriented studies have explored how MCP can be integrated into existing systems to improve efficiency and overall system performance \cite{xu2025towards}. Third, security research has examined attacks against MCP-enabled systems and proposed corresponding mitigation frameworks \cite{mcpSecurity2025}.

While these studies have advanced the technical understanding of MCP, they still provide limited insight into how MCP is actually used in enterprise environments. In particular, the current literature offers little empirical evidence on \textit{where and why enterprises adopt MCP}, \textit{how it is embedded into real LLM-based workflows}, and \textit{what concrete challenges emerge during deployment, coordination, and maintenance}. As a result, existing research has not yet provided a clear picture of MCP's practical role in supporting cross-system invocation, multi-tool collaboration, and enterprise task orchestration. More importantly, current work remains limited in explaining the \textbf{practical bottlenecks} that shape enterprise adoption. These bottlenecks may include integration complexity across heterogeneous systems, coordination frictions across MCP components and teams, difficulties in fault localization and recovery, and the lack of unified standards for scalable deployment.

A further limitation is that the enterprise-facing \textbf{security and privacy implications} of MCP remain insufficiently understood. Existing studies have begun to discuss MCP-related threats, but we still know relatively little about how practitioners perceive security risks in operational settings, what kinds of risks are most salient across industries, and how these concerns compare with alternative approaches such as LLMs using function calling alone. This gap is especially important because security, privacy, and governance concerns directly affect whether MCP can be sustainably deployed in enterprise contexts \cite{narajala2025enterprise,hou2025model}.

These limitations have created a disconnect between MCP's technical development and its enterprise adoption. To narrow this gap, it is necessary to obtain direct insights from enterprise practitioners. Such evidence can help identify the core application scenarios, major deployment challenges, security and privacy concerns, and future expectations surrounding MCP adoption \cite{mai2019mcp}. It can also provide a stronger empirical basis for future research and support the development of MCP solutions that are better aligned with industrial needs and deployment realities \cite{errico2025securing}.

To address these gaps, this study investigates enterprise perspectives on MCP through semi-structured interviews with 20 practitioners. The participants cover multiple MCP-related roles and come from eight companies across two broad industry domains: the Internet industry and the financial industry. The study focuses on four aspects of enterprise MCP practice, namely its current deployment status, its practical value in LLM-based workflows, the challenges and risks arising during adoption, and practitioners' expectations for future improvement. Based on these insights, this study makes three main contributions:

\begin{itemize}
    \item It offers an early empirical understanding of MCP adoption and use in enterprise settings.

    \item It reveals the key challenges, security concerns, and practitioner expectations surrounding MCP deployment.

    \item It derives practical insights for the future improvement of MCP in terms of standardization, usability, and enterprise readiness.
\end{itemize}
\section{Related Work and Background}
This section reviews the evolution of LLM tool integration, the core definition and cutting-edge research progress of the MCP, and the research gaps that motivate this study.

\subsection{Evolution of LLM Tool Integration}\label{AA}
%The integration of LLMs with external tools has undergone continuous optimization centered on collaboration efficiency and ecological compatibility, going through three critical phases that address the core limitations of the preceding stage in turn:
The integration of LLMs with external tools has evolved around improving collaboration efficiency and ecosystem compatibility, progressing through three key phases, each addressing the core limitations of the previous one \cite {liu2024toolace}.

\textbf{Early-stage single-tool function calling.} Pioneered by the OpenAI function calling API \cite{bouchard2024building}, this phase enabled LLMs to perform structured invocations of isolated tools such as calculators and search engines\cite{zhang2024large}. However, it lacked cross-tool context persistence, requiring the retransmission of task metadata for each invocation\cite{kim2024llm} and thus failing to support complex multi-step workflows\cite{wang2025function}.

\textbf{Second-stage agent orchestration frameworks.} Frameworks including LangChain\cite{goyal2025introduction} and AutoGPT \cite{chen2023autoagents} introduced task planning and tool selection logic\cite{lin2025mao}, realizing chained invocations of multiple tools. Nevertheless, their reliance on proprietary protocols led to ecological silos, resulting in high migration costs when switching between different LLMs or tool ecosystems\cite{haddad2017connecting}.

\textbf{Current-stage standardized middleware protocols.} The MCP\cite{anthropic_mcp_2024} emerged as a standardized middle-layer protocol, which breaks down platform and ecological barriers by defining a unified interface for context sharing and tool interaction\cite{hou2025model}. It eliminates vendor lock-in and enables seamless collaboration across heterogeneous systems, thereby meeting the requirements of large-scale deployment.

\subsection{Model Context Protocol}\label{AA}
MCP is a standardized middle-layer protocol designed for the collaboration between LLMs and tool ecosystems. Its core objective is to tackle the issues of context fragmentation, non-standardized interaction, and high architectural coupling in the integration of LLMs with multiple tools and systems, thereby enabling efficient, flexible, and secure collaboration. MCP tool utilisation features four core components (Fig.~\ref{fig:Four components of the MCP protocol}): Host, Client, Server, and Resource (e.g., database) \cite{yang2025survey}, which form the fundamental architectural pillars for its practical implementation. Its core technical characteristics are as follows:

\begin{figure}[htbp]
    \centering
    \includegraphics[width=\linewidth]{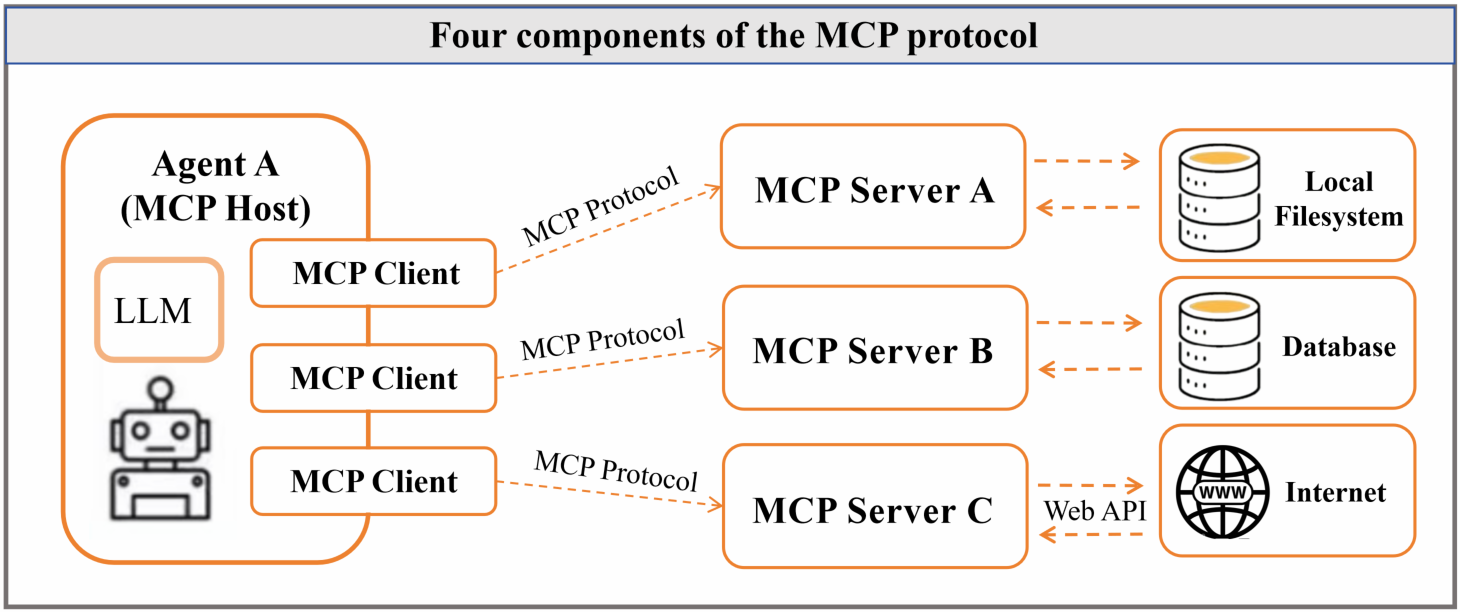}
    \caption{Four Components of the MCP Protocol}
    \label{fig:Four components of the MCP protocol}
\end{figure}

 %\textit{Unified Context Abstraction and Encapsulation.} It defines a standardized structured format to encapsulate task context information including user preferences, tool execution history and security policies in a unified manner, achieving standardized transmission and parsing of information and avoiding incompatibility in interaction formats.

\textbf{Unified context abstraction and encapsulation.} MCP defines a standardized, structured format for representing task context, including user preferences, tool execution history, and security policies, enabling consistent transmission and parsing of contextual information and avoiding incompatibilities in interaction formats.

\textbf{Decoupled layered architecture.} As an independent middle layer, MCP decouples the LLM layer from the tool layer. Both layers can evolve independently against a unified interface, which substantially improves system scalability and maintainability and better accommodates the dynamic evolution of enterprise technical architectures.

% \textbf{Decoupled Layered Architecture: }As an independent middle layer, it achieves complete decoupling between the LLM layer and the tool layer. Both layers can be iterated independently based on a unified interface, which greatly improves the scalability and maintainability of the system and adapts to the dynamic evolution of enterprise technical architectures.

%\textbf{Distributed Dynamic Context Management: }It supports real-time context update, sharing and fine-grained access control across multiple LLM instances and various types of tools, ensuring the consistency, security and availability of context in distributed environments.

\textbf{Distributed dynamic context management.} MCP supports real-time context updates, sharing, and fine-grained access control across multiple LLM instances and heterogeneous tools, helping ensure the consistency, security, and availability of context in distributed environments.

\subsection{Current Research Status of MCP}\label{AA}

\textbf{MCP benchmarks.} Related research\cite{luo2025evaluation,wang2025mcp,yang2025mcpsecbench,liu2025mcpeval} introduces MCP benchmarks that adopt MCP servers, evaluators, and open-source extensible frameworks to evaluate LLMs in real-world tasks, and addresses the limitations of existing MCP benchmarks with this design.\cite{mo2025livemcpbench}.
% 太少了 每个要加至少3-5个

\textbf{MCP integration and performance optimization.} This part of the research\cite{xu2025towards,li2025energyplus,da2025beyond} mainly focuses on how to incorporate MCP into their existing systems to achieve performance optimization.
% 加ref 
% \textbf{Security and Compliance: }Existing research has analyzed the types of attacks targeting MCP integrated systems \cite{mcpSecurity2025} and proposed corresponding security frameworks to mitigate such threats, and has also explored basic protection schemes including role-based access control and audit logging for MCP deployment\cite{fu2025mcp}. Yet these measures only involve passive defense, without designing active defense mechanisms against core risks such as prompt injection and context tampering. In addition, customized solutions tailored to industry compliance requirements have not been developed.

\textbf{Security and compliance.} Research on MCP security has recently emerged as a critical and thriving topic \cite{gan2024navigating,hasan2025model,zhao2025mcp,deng2025ai}. Specifically, \cite{yang2025mcpsecbench} analyzed 17 types of attacks targeting MCP-integrated systems, proposed security frameworks to mitigate these threats \cite{mcpSecurity2025}, and put forward baseline protection measures, such as role-based access control (RBAC) and audit logging for MCP deployment scenarios \cite{fu2025mcp}. However, these protective measures primarily adopt a passive defense approach; they lack proactive mechanisms to guard against core security risks (e.g., prompt injection and context tampering) and fail to provide customized solutions tailored to the compliance requirements of specific industries.

\subsection{Research Gaps and Research Motivation}\label{AA}

Although existing studies have made initial progress in MCP benchmarks, MCP integration, and security compliance, they still exhibit clear limitations that hinder the practical deployment and standardized development of MCP in real enterprises.

First, existing MCP benchmark research mainly focuses on evaluation framework design and tool invocation testing\cite{luo2025evaluation,wang2025mcp,yang2025mcpsecbench,liu2025mcpeval}, but lacks systematic empirical evidence from industrial deployment scenarios. Most benchmarks are constructed in laboratory environments, making it difficult to reflect the actual operation status, pain points, and requirements of MCP in large-scale enterprise systems.

Second, research on MCP integration and performance improvement only explores how to embed MCP into existing systems \cite{xu2025towards,li2025energyplus,da2025beyond}, but fails to systematically summarize the practical challenges encountered during enterprise deployment, such as architectural complexity, cross-team collaboration, state management, and industry-specific adaptation.

Third, research on MCP security and compliance primarily concentrates on attack discovery and passive defense mechanisms\cite{gan2024navigating,hasan2025model,zhao2025mcp,deng2025ai}, while lacking proactive protection schemes and customized compliance solutions for different industrial scenarios.

Above all, the existing literature remains at the technical design and experimental level, and there is a lack of empirical research based on real enterprise practices. This leads to an obvious gap between academic research and industrial applications, and enterprises lack reliable guidance for MCP adoption, technical selection, and risk control.

To fill these gaps, this research conducts in-depth interviews with 20 industrial participants involved in MCP deployment, aiming to systematically explore the practical application status, core challenges, and user expectations of MCP in real-world scenarios. The findings of this work are expected to provide empirical support and practical guidance for the technical iteration, standardized construction, security enhancement, and large-scale promotion of MCP.

%kh 已改   针对这个建议“ 这段要重改。要和上面的MCP research连接起来。可以说section II-C 的research对应的limitation是啥，因此我们做了这个work来干啥。不然下面几个点没有任何对应。”
\section{RESEARCH METHODOLOGY}

\begin{figure}[htbp]
    \centering
    \includegraphics[width=\linewidth]{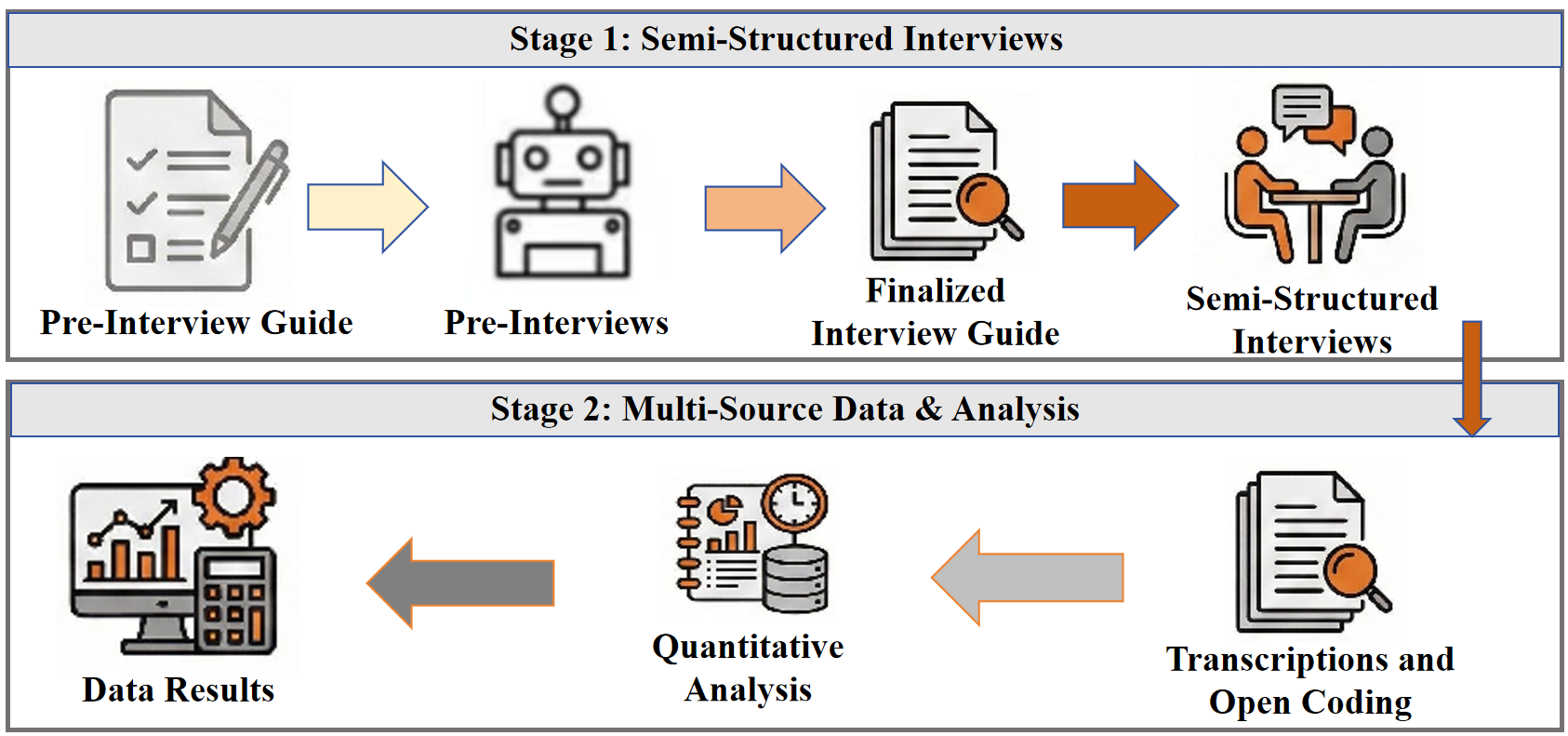}
    \caption{The Overview of the Research Methodology}
    \label{fig:research_methodology}
\end{figure}
% ========== 插入结束 ==========
\subsection{Research Methods}\label{AA}

Our research adopts a semi-structured interview approach\cite{adeoye2021research}, an overview of which is presented in Fig.~\ref{fig:research_methodology}. We use expert sampling, a subtype of purposive sampling \cite{baltes2022sampling}, and conduct semi-structured interviews with 20 participants in the MCP field, preceded by 3 pre-interviews to refine the interview guide.

\textit{(a)} \textit{Interview Outline Design:} Initially, we formulated preliminary questions based on research experience and references to other literature\cite{ma2025practitioners,bingham1931interview,nsengiyumva2026perceptions}, and then verified them through exploratory interviews with 3 participants who were not included in the 20 participants in the formal interviews. %kh
% 这里要说 这三个人不参与访谈（不在那20人）。然后大概介绍一下3个人是什么背景。还有注意区分practitioner和participant。后面好几次写错了。√
Based on the insights from the pre-interviews, we developed a semi-structured interview guide consisting of open-ended questions \cite{le2014loaded}. To reduce response bias, we allowed participants to answer ``I don’t know'' when they were unfamiliar with or unclear about the questions.
% 这里不对，这个流程是survey的，interview的问题都是问答题吧？不提供我不知道的选项。
This preliminary process eventually formed 31 core questions, which remained consistent in subsequent formal interviews. 
The following is a detailed explanation of the workflow and user scenarios of our interview design. We have uploaded the interview guide online\cite{Anonymity2026}.

\textbf{1) }\textbf{Demographics.} This section collects demographic data and the composition of the interviewed participants, including their countries of residence, primary work roles, professional experience, programming languages, and team sizes.

\textbf{2) }\textbf{MCP practices and issues.} First, the participants will be asked if they have experience in developing MCP, and then the question ``Which of the following roles is closest to your current identity in the process of using MCP?'' will be raised, such as ``MCP Host'', ``MCP Client'', ``MCP Server'', ``MCP User'' \cite{yang2025survey}. Then, the participants will be asked questions like ``Please share an example of your use of MCP'' and ``How does MCP fit into your existing workflow?'' And questions about the problems encountered will also be asked, for example, ``What was the biggest obstacle you encountered in the process of using the MCP?''

\textbf{3) }\textbf{MCP technical capability assessment.} 
% In this section, detailed questions are mainly asked for technical assessment, divided into four aspects:
% Irreplaceability Verification: ``Please share an example of a task that MCP can solve but a pure LLM cannot'' Through these questions, the irreplaceability of MCP is analyzed.
% Technical Debt Comparison: Compare the differences between using LLMs and function calls, and between using LLMs and MCP, across five dimensions: architectural complexity, operational cost and maintenance, and development efficiency and scalability. 
% In addition, by asking ``What changes have occurred in the following aspects after introducing MCP compared to using the best pure LLM?''.
% Technical Fault Analysis: Questions such as ``Have you ever encountered any faults when using MCP?'' will be asked. The participants are required to provide details such as the fault manifestations, average fault repair time, and factors hindering rapid repair.
% % participants不大写（全文检查）
% Indicator Assessment: By asking ``Which indicators do you think should be used to evaluate the performance and recovery ability of MCP?'', understand which aspects of MCP the participants value most.
% 这里要写成一个段落的形式，前后要连接起来，几个方面可以用斜体或者粗体表明。然后每一个方面要说，通过这个问题，我们要了解interviewee xxx的具体的情况。这段要重改。
We conduct technical evaluations through detailed questions, which are divided into four dimensions. \textit{Irreplaceability Verification:} By asking the question “Please share an example of a task that MCP can solve but a pure large language model cannot”, we aim to understand the participants’ real experience of the irreplaceable value of MCP in practical scenarios. \textit{Technical Comparison:} We compare the differences between “LLM + function calling” and “LLM + MCP” from five dimensions: architectural complexity, operational cost, maintenance cost, development efficiency, and scalability. We also ask “What changes have occurred in the following aspects after introducing MCP compared with using the pure LLM?” to understand the actual improvements and technical differences perceived by the participants. \textit{Technical Fault Analysis:} By asking questions such as “Have you ever encountered any faults when using MCP?”, we require the participants to describe the fault manifestations, average fault repair time, and factors hindering rapid repair, so as to understand the actual faults and difficulties they encountered in the use and operation of MCP. \textit{Indicator Assessment:} By asking the question “Which indicators do you think should be used to evaluate the performance and recovery ability of MCP?”, we aim to understand which technical indicators and evaluation dimensions of MCP the participants value most.%kh

\textbf{4) }\textbf{Security and privacy.} This section investigates whether participants in the Internet industry and FinTech industry have ``encountered risks of data leakage or attacks when using MCP'' and ``what security and privacy technologies are used when using MCP''. For the client and host roles, the questions are: ``Are you clear about the logic of MCP's model selection? Can you intervene in incorrect scheduling?'', ``Where do you think a fault would be most dangerous?'' 

\textbf{5) }\textbf{Future expectations.} This section focuses on interviewing participants to summarize the pain points in using MCP and their imagination and expectations for the future usage scenarios of MCP, so as to obtain the gap between the current usage status of MCP and the future expectations. The following questions are asked: ``What are the bottleneck links in current daily use?'', ``What kind of MCP experience do you hope to improve the most?'', ``What future trends of MCP do you predict?''

\textbf{6) }\textbf{Counterintuitive insights.} At the end of the questionnaire, a question about counterintuitive insights is raised to better capture the cognition that is easily taken for granted as correct in daily MCP development but is actually full of pitfalls. The question is: ``Which discovery completely subverted your expectations in MCP practice?''

\textit{(b)} \textit{Implementation Process:}
\begin{itemize}
 \item\textbf{Step 1. Pre-interview:} Conduct pre-interviews with 3 participants and adjust the language expression of the outline (e.g., change the question from ``\textit{Compare LLM and MCP}'' to ``\textit{Compare LLM with function call and LLM with MCP}'' to reduce users' understanding cost). 
\item \textbf{Step 2. Formal interview:} Adopt the online video form, with each session lasting 90--120 minutes. Record the whole process and transcribe it into text (collect approximately 400,000 words of original data in total).
\item \textbf{Step 3. Supplementary interview:} Conduct secondary interviews with 10 key participants (each session lasts 30 minutes) to clarify vague views.\end{itemize}

% After completing 20 formal interviews, we stopped inviting additional participants to ensure that all participants were professional MCP participants and that the data had reached saturation no new important insights emerged, indicating that the relevant research dimensions had been fully covered
% 针对声称在进行20次访谈后达到了饱和状态，这看似合理，但论文应描述如何评估饱和状态（例如，饱和网格、编码本稳定性），而不是仅仅断言这这个评价

We adopted the principle of theoretical saturation to determine sample adequacy.
Data collection ceased after 20 interviews, as no new concepts, challenges, or categories emerged in successive interviews.
The coding framework was fully stabilized, with consistent themes repeated across participants, indicating that the sample was sufficient to capture the core perspectives in the studied contexts.\cite{ma2025practitioners}.

\subsection{Research Participants}\label{AA}

In this study, 20 professional MCP users and developers were selected, covering 9 from FinTech (e.g., HSBC) and 11 from the Internet (e.g., Huawei).
% 要说具体公司。
%They involve the four core roles of MCP: MCP Host, MCP Client, MCP Server, and MCP User. The details are as follows (TABLE ~\ref{tab:participants}):
% 不能说他们是这些角色呀，这些都是机器！要说他们参与了使用或开发xxx等，这个要写详细，一定要严谨。√
%kh 改 
The use, development, deployment, and management of MCP involve multiple stakeholders, and the entire process relies heavily on MCP's three core functional components: the MCP host, the MCP client, and the MCP server. Specifically, relevant personnel (including Architects, Developers, and Business Analysts) participate in the design and optimization of the MCP host, the configuration and invocation of the MCP client, the deployment and maintenance of the MCP server, and the operation and interaction of MCP users. These core components work collaboratively with the participation and operation of these personnel to support the normal operation of the entire MCP protocol system. The specific functions and participation details of each component are as follows (Table~\ref{tab:participants_all}):%kh

% 导言区需加载的包（缺一不可）
% \usepackage{booktabs}  % 支持三线表
% \usepackage{array}     % 支持p{}列格式
% \usepackage{caption}   % 必须加载此包，才能修改标题分隔符

% 完全保留你设置的列宽
\begin{table*}[htbp]
\footnotesize
\captionsetup{justification=centering, labelsep=none}
\caption{\\PARTICIPANT DEMOGRAPHICS AND EXPERIENCE DISTRIBUTION}
\label{tab:participants_all}
\centering
\renewcommand{\arraystretch}{0.9}
\setlength{\tabcolsep}{3pt}
\begin{tabular}{p{0.04\linewidth} p{0.18\linewidth} p{0.11\linewidth} p{0.16\linewidth} c c c c c c c c}
\toprule
\textbf{ID} & \textbf{Role} & \textbf{Industry} & \textbf{MCP Role} & \textbf{Management} & \multicolumn{3}{c}{\textbf{Experience}} & \multicolumn{4}{c}{\textbf{Team Size (ppl.)}} \\
\cmidrule(lr){6-8} \cmidrule(lr){9-12}
& & & & & \textbf{$<$3y} & \textbf{3-10y} & \textbf{$>$10y} & \textbf{6-20} & \textbf{21-40} & \textbf{41-80} & \textbf{$>$80} \\
\midrule
P1  & Senior Architect      & FinTech   & Host/Client/Server  & Yes  & 0 & 0 & 1 & 0 & 0 & 0 & 1 \\
P2  & Senior Architect      & Internet  & Host/Client/Server  & Yes  & 0 & 0 & 1 & 0 & 0 & 0 & 1 \\
P3  & Senior Architect      & Internet  & Host/Client/Server  & Yes  & 0 & 0 & 1 & 0 & 0 & 0 & 1 \\
P4  & Senior Architect      & Internet  & Host/Client/Server  & Yes  & 0 & 1 & 0 & 0 & 1 & 0 & 0 \\
\midrule
P5  & Software Developer    & FinTech   & Client/Server       & No   & 0 & 1 & 0 & 0 & 1 & 0 & 0 \\
P6  & Software Developer    & Internet  & Client/Server       & No   & 0 & 1 & 0 & 0 & 1 & 0 & 0 \\
P7  & Software Developer    & FinTech   & Client/Server       & No   & 0 & 1 & 0 & 0 & 1 & 0 & 0 \\
P8  & Software Developer    & Internet  & Client/Server       & No   & 0 & 1 & 0 & 0 & 1 & 0 & 0 \\
P9  & Software Developer    & FinTech   & Client/Server       & No   & 0 & 1 & 0 & 0 & 1 & 0 & 0 \\
P10 & Software Developer    & Internet  & Client/Server       & No   & 0 & 1 & 0 & 0 & 1 & 0 & 0 \\
P11 & Software Developer    & FinTech   & Client/Server       & No   & 0 & 1 & 0 & 0 & 0 & 1 & 0 \\
P12 & Software Developer    & Internet  & Client/Server       & No   & 0 & 1 & 0 & 0 & 0 & 1 & 0 \\
P13 & Software Developer    & FinTech   & Server              & No   & 0 & 1 & 0 & 0 & 0 & 1 & 0 \\
P14 & Software Developer    & Internet  & Server              & No   & 1 & 0 & 0 & 0 & 0 & 1 & 0 \\
\midrule
P15 & Business Analyst      & FinTech   & User                & No   & 0 & 1 & 0 & 0 & 1 & 0 & 0 \\
P16 & Business Analyst      & Internet  & User                & No   & 0 & 1 & 0 & 0 & 1 & 0 & 0 \\
P17 & Business Analyst      & FinTech   & User                & No   & 0 & 1 & 0 & 0 & 1 & 0 & 0 \\
P18 & Business Analyst      & Internet  & User                & No   & 0 & 1 & 0 & 0 & 0 & 1 & 0 \\
\midrule
P19 & Algorithm Designer    & FinTech   & Server              & Yes  & 0 & 1 & 0 & 0 & 1 & 0 & 0 \\
P20 & Algorithm Designer    & Internet  & Server              & Yes  & 0 & 0 & 1 & 0 & 0 & 0 & 1 \\
\midrule
\textbf{Total} & - & \textbf{9F, 11I} & - & \textbf{6Yes,14No} & \textbf{1} & \textbf{15} & \textbf{4} & \textbf{0} & \textbf{11} & \textbf{6} & \textbf{3} \\
\bottomrule
\end{tabular}
\end{table*}

% %Sample supplementary note: 
% Tables ~\ref{tab:mcp_roles} and ~\ref{tab:practitioner_experience} summarize the background of the interviewed interview participants, where $\#$ denotes the number of interviewees. All participants have 2--20
% years of work experience. Among them, 15 people (75\%)
% have 3--10 years of experience (core execution layer), 3
% people (15\%) have 10--15 years of experience (technical
% management layer), 1 people (5\%) have 20 years of experience
% (enterprise decision makers), and only 1 person (5\%) has 1--3 years of experience (newcomer’s perspective). The team
% size ranges from 6--160 people (large-scale enterprises),
% ensuring that the sample has multiple perspectives across industries and covers all core roles of MCP (technical execution,
% management, decision-makers, and end-users). An overview of the distribution of participants’ roles and their experiences is shown in Table~\ref{tab:practitioner_experience}.
% %修改了根据新表把#去掉了

%Sample supplementary note: 
Tables ~\ref{tab:participants_all} summarize the background of the interviewed interview participants; all participants have 2--20
years of work experience. Among them, 15 people (75\%)
have 3--10 years of experience (core execution layer), 3
people (15\%) have 10--15 years of experience (technical
management layer), 1 people (5\%) have 20 years of experience
(enterprise decision makers), and only 1 person (5\%) has 1--3 years of experience (newcomer’s perspective). The team
size ranges from 6--160 people (large-scale enterprises),
ensuring that the sample has multiple perspectives across industries and covers all core roles of MCP (technical execution,
management, decision-makers, and end-users). An overview of the distribution of participants’ roles and their experiences is shown in Table~\ref{tab:participants_all}.
%kh 加解释done
% 要解释呀。

\subsection{Research Questions}
Based on the participants' input, we formulate the following four Research Questions (RQs):

\textbf{RQ1: What is the current status of MCP's practical application, and what issues does it face?} On the one hand, to understand how MCP is actually being used, we examine specific application scenarios reported by the interviewees. For example, which complex tasks MCP is used to help LLMs accomplish, and the scope and depth of its adoption across different domains. On the other hand, we focus on identifying the issues and challenges encountered during the real-world use and development of MCP, including (but not limited to) technical limitations, high development and maintenance costs, poor compatibility with other tools, and usability problems in day-to-day operations. Our findings indicate that MCP has been widely deployed in scenarios where LLMs collaborate with multiple systems, and its core value lies in enabling language models to access external tools, files, databases, and contextual resources in a standardized and secure manner. At the same time, MCP still faces prominent pain points, particularly in cross-component collaboration and compatibility with certain mainstream tools.

\textbf{RQ2: How important is MCP to participants, and how does it influence their work?} This question first examines how participants who use MCP in their daily work perceive its importance, and then investigates its concrete impact on their workflows, efficiency, and working patterns. Our findings indicate that all participants (100\%) regarded MCP as critical to their work. 
In terms of concrete benefits, 80\% reported that MCP significantly improves their efficiency by simplifying the connection between LLMs and external tools; 75\% stated that it optimizes their working patterns (e.g., internal task management platforms), by standardizing task execution and collaboration processes; and 70\% noted that it lowers the technical barrier for operation and development, thereby reducing the difficulty of related tasks.

\textbf{RQ3: What security and privacy risks arise in MCP-based systems, and how do LLM+MCP architectures differ from LLMs using only function calling?} 
This question first examines the security and privacy risks introduced by MCP-based systems and then compares LLM+MCP architectures with LLMs that rely solely on function calling. Our findings show that, in terms of risk, participants from Internet and FinTech companies emphasized different concerns: Internet teams focused on integration robustness and data leakage, whereas FinTech teams were more concerned with regulatory compliance, access control, and auditability. For the architectural comparison, 80\% of participants reported that although LLM+MCP architectures are slightly more complex than LLMs with function calling, they require less long-term development effort and provide better maintainability.

\textbf{RQ4: What are participants' expectations for MCP, and what challenges occur in multi-model collaboration?} This question explores the desired future capabilities of MCP and observations from multi-model (or multi-agent) usage scenarios. Participants expect MCP to move toward stronger protocol standardization, the creation of a unified tool marketplace, and more powerful low-code integration capabilities, alongside enhanced support for collaborative operations and maintenance. They also highlight the need for more systematic security and privacy protections, especially in multi-model and multi-tool usage scenarios, where organizations must collaborate under joint governance and policy constraints.

\subsection{Data Analysis Method}
A mixed research method, incorporating both qualitative and quantitative research, is adopted to combine in-depth insight with empirical data support\cite{ostlund2011combining}. 
First, we transcribed the interview content verbatim and
used open coding\cite{spencer2009card, ma2025practitioners} to generate an initial code set. 
Subsequently, the first two authors for cross-validation and
optimization of the coding logic. Cohen’s kappa coefficient reached 0.89 (high consistency). Regarding classification disagreements, the team reached a consensus through in-depth discussions. Based on a rigorous process, we extracted 6 core issues (RQ1) during the use of MCP from the interviews and 3 aspects of key improvement expectations (RQ4) from users for this tool.
% 什么是log mcp？√
% Quantitative analysis \cite{nevin1984quantitative} focuses on interview data and provides empirical support for research conclusions through statistical modeling and visualization methods. Specifically, we conducted statistical analysis on quantifiable content extracted from semi-structured interviews. All quantifiable data were visually presented through statistical charts to enhance the readability and persuasiveness of the analysis results. For the open-ended, exploratory content of the semi-structured interviews, we adopted qualitative analysis to deeply mine the hidden views, demands and practical experiences of the participants.
Quantitative analysis \cite{nevin1984quantitative} was applied to the interview data to provide empirical support for the research findings through statistical modeling and visualization. Specifically, we analyzed quantifiable content extracted from the semi-structured interviews and presented it using statistical charts to improve the clarity and persuasiveness of the results. For the open-ended, exploratory responses, qualitative analysis was employed to uncover participants’ underlying views, needs, and practical experiences.
\section{Results}
The interviewees had 1--20 years of professional experience, worked in teams of 6--160 members, and mastered full-stack development languages including Java, Python, C++, and Rust. To comply with academic ethics and data privacy regulations, all specific enterprise names involved in the study were anonymized. Interview data were analyzed via open coding. Combined with empirical cases extracted from meeting minutes, the analysis yielded systematic research outcomes.

\subsection{RQ1: What is the current status of MCP's practical application, and what issues does it face?}\label{AA}
This research question aims to investigate participants’ experiences in using MCP tools, focusing on daily usage scenarios, understanding and application of MCP, as well as challenges encountered during tool deployment.

\textit{1)} \textit{Current status of Model Context Protocol:}
MCP has been widely adopted across the FinTech and Internet sectors, exhibiting distinct industry-specific application characteristics driven by compliance requirements and business scenario differences, and has formed a standardized role-based usage model.
The FinTech industry prioritizes internal development efficiency and deterministic task execution. To meet data privacy and regulatory compliance mandates, it completely avoids external API connections. Core application scenarios include team task management, internal tool integration, and integration with internal requirements analysis systems. According to participant feedback, MCP implementation was perceived to substantially shorten cross-team development cycles and reduce error rates in complex task execution. The Internet industry regards MCP as the core infrastructure for integrating internal and external tools, focusing on AI-assisted coding, cross-platform data processing (e.g., PDF structured extraction, user behavior data analysis), and multi-system collaboration (e.g., automated test case review, cross-platform message synchronization). As reported by Internet industry practitioners, MCP is used daily to complete work tasks, with its usage frequency deeply integrated into core business processes. As participants noted: \textit{``Practical cases of the team using MCP mainly involve extracting Markdown from PDFs and structuring the data for LM to call. It's worth noting that the data formats they handle are diverse and special, including payment data such as XML and MX''}(Participant 16).
.

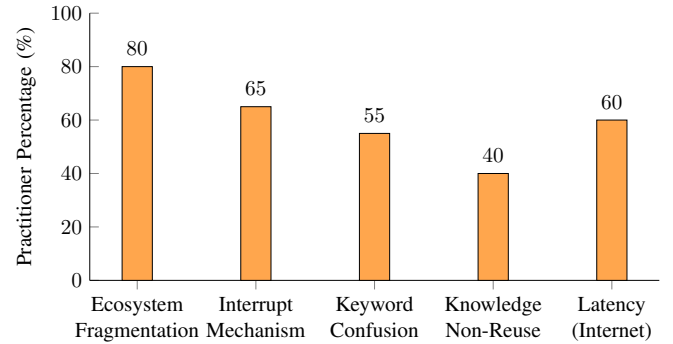
\begin{figure}[htbp]
  \centering
  \begin{tikzpicture}[scale=0.8, transform shape]
    \begin{axis}[
      width=11cm, height=6cm,
      ybar,
      ymin=0, ymax=100,
      ylabel={Practitioner Percentage (\%)},
      % 统一：10pt字体 + 12pt行间距（行业标准）
      ylabel style={font=\fontsize{10}{12}\selectfont, align=center}, 
      xlabel={},
      xlabel style={font=\fontsize{10}{12}\selectfont, align=center, yshift=-20pt},
      xtick={1,2,3,4,5},
      xticklabels={
        Ecosystem\\Fragmentation, 
        Interrupt\\Mechanism, 
        Keyword\\Confusion, 
        Knowledge\\Non-Reuse ,
        Latency\\(Internet)
      },
      xticklabel style={
        font=\fontsize{10}{12}\selectfont, 
        align=center,
        yshift=0pt 
      },
      bar width=0.5cm,
      nodes near coords, 
      nodes near coords align={center},
      every node near coord/.style={
        % 修正：从10/8改为10/12，字体大小保持10pt，行间距统一
        font=\fontsize{10}{12}\selectfont,
        yshift=0.3cm
      },
      % 统一：10pt字体 + 12pt行间距
      tick label style={font=\fontsize{10}{12}\selectfont, align=center},
      label style={font=\fontsize{10}{12}\selectfont, align=center},
      grid=minor,
      minor grid style={dashed, gray!30},
      axis lines*=left,
      enlarge x limits=0.1,
    ]
      \addplot [fill=orange!70] coordinates {(1,80) (2,65) (3,55) (4,40) (5,60)};
    \end{axis}
  \end{tikzpicture}
  \caption{Key Technical Bottlenecks of MCP Adoption}
  \label{fig:technical_bottlenecks}
  \vspace{-6pt}
\end{figure}

\textit{2)} \textit{Issues of Model Context Protocol:} % Participants in both industries face common bottlenecks and industry-specific pain points when using MCP, with core challenges summarized (Fig.~\ref{fig:technical_bottlenecks} ). MCP lacks unified industry standards and interface specifications, resulting in incompatibility between different Agent frameworks (e.g., LangGraph, ADK). 16 (80\%) of surveyed participants reported that their projects need to invest additional resources in developing adapters, among whom 40\% indicated that adapter development consumes more than 20\% of total project development resources. Cross-component collaboration issues are prominent in both industries: 13 (65\%) of participants stated that MCP in distributed architectures lacks a native interrupt mechanism, failing to support flexible workflows such as ``pause-manual interaction-resume execution''\cite{captheorem1999}. As participants noted: \textit{``When MCP executes tasks sequentially (e.g., Tasks 1-10), it may need to interact with LLMs to obtain information when running Task 5, however, after receiving LLM feedback, MCP may restart execution from Task 1''}. Such issues increase development complexity by 30\%. 11 (55\%) of surveyed participants across both industries reported that MCP relies on keyword matching for tool scheduling, leading to frequent ``recognition confusion'' where the target tool cannot be invoked accurately. Some participants commented: \textit{``MCP interfaces of different frameworks vary significantly in parameter formats, requiring re-adaptation for each integration and consuming substantial development time''}; \textit{``Keyword-based scheduling often triggers irrelevant tools by mistake, which instead reduces work efficiency''}. 修改强制量化的表达，并且加上了受访者编号，finding1 里面的不合适量化表达也删除了，原来的是Finding 1. Over 80% of participants are plagued by MCP ecosystem fragmentation and collaboration barriers, which stand as the core challenges.
Participants in both industries face common bottlenecks and industry-specific pain points when using MCP, with core challenges summarized (Fig.~\ref{fig:technical_bottlenecks}).
MCP lacks unified industry standards and interface specifications, resulting in incompatibility between different Agent frameworks (e.g., LangGraph, ADK).
Many participants reported that their projects needed to invest additional resources in developing adapters, with several noting that adapter development occupied a substantial portion of overall development effort. As participants noted: \textit{`` The MCP is mainly responsible for tool hosting, but it needs to be combined with the Agent Orchestration Layer (such as frameworks like ADK and Lang) to realize its value. These frameworks each define tool standards, resulting in the MCP having to adapt to different ecosystems''}(Participant 11).
.

Cross-component collaboration issues are prominent in both industries.
Many participants stated that MCP in distributed architectures lacks a native interrupt mechanism, failing to support flexible workflows such as ``pause-manual interaction-resume execution''.
As participants noted: \textit{``When MCP executes tasks sequentially (e.g., Tasks 1-10), it may need to interact with LLMs to obtain information when running Task 5; however, after receiving LLM feedback, MCP may restart execution from Task 1''}(Participant 6).
Such issues were widely perceived to increase development complexity.

Many participants across both industries reported that MCP relies on keyword matching for tool scheduling, leading to frequent ``recognition confusion'' where the target tool cannot be invoked accurately.
Some participants commented:
\textit{``MCP interfaces of different frameworks vary significantly in parameter formats, requiring re-adaptation for each integration and consuming substantial development time''}(Participant 15);
\textit{``Keyword-based scheduling often triggers irrelevant tools by mistake, which instead reduces work efficiency''}(Participant 3).
\begin{center}
\vspace{-8pt}
    \resizebox{\linewidth}{!}{
\begin{tabular}{l!{\vrule width 1pt}p{0.9\columnwidth}}
    \makecell{{\LARGE \faLightbulbO}}  &\textbf{Finding 1.} 
    A large majority of participants are plagued by MCP ecosystem fragmentation and collaboration barriers, which stand as the core challenges. 
\end{tabular}}
\vspace{8pt}
\end{center} 

100\% of participants indicated that quickly locating the problem when encountering a malfunction is the most significant obstacle to troubleshooting. They also expressed a desire for automated detection tools in the future to help them better pinpoint issues. As one practitioner noted: \textit{``If there are bugs in the MCP library itself, it will be quite troublesome to fix them. The maintenance process of open-source libraries may not be efficient enough. The proper approach is to submit an issue rather than make modifications on your own''}(Participant 7). As participants noted: \textit{`` The new MCP protocol is similar to the initial growing pains of the "Type-C" interface. Many internal tools (such as ServiceNow and DevOps pipelines) have not yet adapted to it, resulting in compatibility issues when connecting with legacy systems. Since the MCP Server has not undergone specialized optimization training or the interpretation of instructions for specific tasks is not clear enough, there are often problems of chaotic command, that is, the large model cannot accurately understand and call the target MCP''}(Participant 13).

\begin{center}
\vspace{-8pt}
    \resizebox{\linewidth}{!}{
\begin{tabular}{l!{\vrule width 1pt}p{0.9\columnwidth}}
    \makecell{{\LARGE \faLightbulbO}}  &\textbf{Finding 2.} 
    100\% of the participants thought that the biggest problem they encountered when repairing MCP faults was the difficulty in locating where the problem occurred and the lack of standardized fault diagnosis tools. 
\end{tabular}}
\vspace{-5pt}
\end{center}

\subsection{RQ2: How important is MCP to participants, and how does it influence their work?}\label{AA}
This research question explores the importance of MCP to participants and its specific impacts on work efficiency, workflows, skills requirements, and collaboration methods.

MCP delivers high value in both industries, with 100\% of participants considering it either critical or beneficial to their work. FinTech participants view it as a key tool for improving internal efficiency. As participants noted: \textit{`` Traditional methods require manually writing a lot of code to handle exceptions, while MCP can simplify this process''}(Participant 3). It solves the problems of pure LLM not working in internal task management systems and the uncertainty in executing complex tasks. Internet industry participants regard it as the core infrastructure connecting LLMs with internal and external tools, breaking through the capability limitations of standalone LLMs to enable cross-system collaboration and process automation. As one practitioner noted: \textit{``The field length conversion scenario must rely on the M toolCP for implementation because the LLM cannot accurately handle such numerical boundary issues''}(Participant 4).

\begin{center}
\vspace{-12pt}
    \resizebox{\linewidth}{!}{
\begin{tabular}{l!{\vrule width 1pt}p{0.9\columnwidth}}
    \makecell{{\LARGE \faLightbulbO}}  &\textbf{Finding 3.} In the Internet and fintech, the obstacles faced when using MCP vary according to enterprise characteristics and compliance restrictions.
\end{tabular}}
%\vspace{-8pt}
\end{center}

% MCP increases complex task processing efficiency by over 40%: FinTech teams shorten cross-team collaboration cycles
% through MCP tool integration and knowledge reuse; Internet
% teams automate data processing and multi-system collabora-
% tion, reducing manual coding and data migration costs. How-
% ever, adoption barriers vary by industry: Internet participants
% face challenges in external tool integration and performance
% optimization, while FinTech participants struggle with knowl-
% edge reuse system construction. As one practitioner noted:
% “MCP has transformed cross-system data synchronization
% from manual to automated operations, significantly improving
% efficiency. However, compatibility and security issues during
% external tool integration often trouble us”. 把第一句删掉了，加了参与者

FinTech teams shorten cross-team collaboration cycles through MCP tool integration and knowledge reuse; Internet teams automate data processing and multi-system collaboration, reducing manual coding and data migration costs. However, adoption barriers vary by industry: Internet participants face challenges in external tool integration and performance optimization, while FinTech participants struggle with knowledge reuse system construction. As one practitioner noted: \textit{``MCP has transformed cross-system data synchronization from manual to automated operations, significantly improving efficiency. However, compatibility and security issues during external tool integration often trouble us''}(Participant 1).

% % \begin{center}
%     \resizebox{\linewidth}{!}{
%     \centering
% \begin{tabular}{l!{\vrule width 1pt}p{0.9\columnwidth}}
%     \makecell{\LARGE \faLightbulbO}  & \textbf{Finding 3.} MCP reshapes industry workflows and practitioner skill requirements.
% \end{tabular}}

% % \end{center}

\begin{center}
\vspace{-12pt}
    \resizebox{\linewidth}{!}{
\begin{tabular}{l!{\vrule width 1pt}p{0.9\columnwidth}}
    \makecell{{\LARGE \faLightbulbO}}  &\textbf{Finding 4.} MCP reshapes industry workflows and practitioner skill requirements, with distinct industry-specific focuses in MCP application(Fig.~\ref{fig:mcp_core_scenarios_f3}).
\end{tabular}}
%\vspace{-8pt}
\end{center} 

\begin{figure}[H]
  \centering
  \begin{tikzpicture}[scale=0.8, transform shape]
    \begin{axis}[
      width=10cm, height=6cm,
      ybar,
      ymin=0, ymax=100,
      ylabel={Practitioner Percentage (\%)},
      % 统一字体：10pt大小 + 12pt行间距
      ylabel style={font=\fontsize{10}{12}\selectfont, align=center}, 
      xlabel={},
      % 核心调整：X轴总标题位置，yshift正数向下移，避免重叠
      xlabel style={font=\fontsize{10}{12}\selectfont, align=center, yshift=-20pt},
      xtick={1,2,3,4},
      xticklabels={
        Internal Task\\Management,
        Automated\\Approval,
        Knowledge\\Reuse,
        External Tool\\Linkage
      },
      % 刻度标签样式统一，复位位置避免偏移
      xticklabel style={
        font=\fontsize{10}{12}\selectfont, 
        align=center,
        yshift=0pt
      },
      bar width=0.5cm,
      nodes near coords, 
      nodes near coords align=center,
      % 统一刻度/标签字体样式
      tick label style={font=\fontsize{10}{12}\selectfont, align=center},
      label style={font=\fontsize{10}{12}\selectfont, align=center},
      grid=minor,
      minor grid style={dashed, gray!30},
      axis lines*=left,
      % 增大X轴外侧间距，适配4个分类
      enlarge x limits=0.15,
      cycle list={orange!80!black, orange!40!white},
      % 图例样式：字体改为10pt，保留原有位置和边框
      legend pos=north east,
      legend style={
        font=\fontsize{10}{12}\selectfont,
        column sep=0.5cm,
        fill=white,
        draw=black,
        inner sep=2pt,
        at={(1.0, 1.15)},
        anchor=north east
      }
    ]
      % 金融科技（深橙）：保留颜色，仅统一数值标签字体
      \addplot [fill=orange!80!black,
        every node near coord/.style={
          font=\fontsize{10}{12}\selectfont,
          yshift=0.2cm, 
          xshift=-0.3cm
        }]
        coordinates {(1,100.00) (2,88.89) (3,77.78) (4,0.00)};
      
      % 互联网（浅橙）：保留颜色，仅统一数值标签字体
      \addplot [fill=orange!40!white, draw=orange!80!black,
        every node near coord/.style={
          font=\fontsize{10}{12}\selectfont,
          yshift=0.2cm, 
          xshift=0.4cm
        }]
        coordinates {(1,27.27) (2,54.55) (3,36.36) (4,72.73)};
      
      \legend{FinTech (9 participants), Internet (11 participants)}
    \end{axis}
  \end{tikzpicture}
 \caption{Core Application Scenarios of MCP}
  \label{fig:mcp_core_scenarios_f3}
  \vspace{-6pt}
\end{figure}
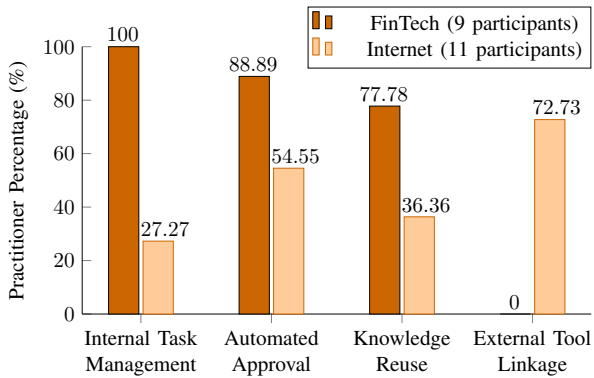

Workflows have been expanded to include MCP tool registration, protocol adaptation, and permission configuration, forming standardized processes from demand initiation to task execution. Collaboration models have shifted toward cross-team and cross-model coordination: Internet teams adopt distributed MCP architectures to avoid single points of failure; FinTech teams break internal barriers through unified tool standards. As participants noted: \textit{`` Global engineers contribute reusable prompts to the promptbook, classify them by technical fields, and then connect them to various agents through MCP''}(Participant 17).

Notably, the two industries exhibit divergent focuses when leveraging MCP, 100\% of FinTech interviewees prioritize MCP for improving daily task management efficiency, while 72.72\% (8 participants) of Internet industry interviewees focus on MCP’s capability to link with external tools. Specifically, FinTech participants use MCP to connect LLMs with internal task management platforms (e.g., Jira/Comfluence), automating team task management and review for efficiency. In contrast, Internet participants rely on MCP to enable LLMs to access the latest real-time data, ensuring the accuracy of LLM-generated answers. As participants noted: \textit{`` Taking the test case review as an example, MCP can directly operate on conference data, while pure LLM requires manual data transfer. Therefore, MCP is very useful in terms of efficiency improvement''}(Participant 17).

Participants are required to master MCP protocols, multi-model adaptation, and fault diagnosis skills, with ``product-oriented thinking'' becoming a core competency for MCP-related technical roles. A participant stated, \textit{``Now, technical development not only requires coding skills but also consideration of MCP tool adaptability and usability, we need to design interfaces from the user’s perspective''}(Participant 7).

\subsection{RQ3: What security and privacy risks arise in MCP-based systems, and how do LLM+MCP architectures differ from LLMs using only function calling?}\label{AA}

This research question focuses on the characteristic differences between MCP-related technical combinations, as well as potential security and privacy risks and mainstream protection measures in MCP deployment, with emphasis on industry-specific features. In two different industries, and even in different departments and different businesses within the same industry, the understanding and application of MCP security measures vary. Through interviews, it is obvious that there are significant differences in the perception of the application of security measures and risk control methods among the interviewees. There are also no relevant security measure guidelines in the industry. As one practitioner noted: \textit{``Occasional failures of the MCP service are caused by the imperfect monitoring and automatic recovery mechanisms, which reflect the instability of the new system in its initial stage. However, the recovery time is usually within ten minutes. The main manifestation of the failure is a response timeout, and the calling process will be directly terminated to avoid output accidents, indicating that the team pays more attention to system security than forcing the completion of the process''}(Participant 14). Some practitioners noted: \textit{``Currently, the team uses permission control and audit logs to ensure security. There isn't much involvement in the use of sensitive words and API encryption at present''}(Participant 7).

%In terms of technical combinations, LLM combined with MCP or LLM combined with function calling exhibits distinct characteristics. Regarding security risks, influenced by external interaction scenarios, the FinTech industry faces almost zero external risks, while the Internet industry confronts diverse risk challenges. 80\% of Internet participants regard permission overstepping and prompt injection as MCP’s core security risks; mainstream protection measures are fine-grained access control, audit logging, and API encryption, with inadequate active defense systems. 100\% of developers confirm MCP+LLM features slightly higher architectural complexity, but lower development difficulty and better long-term maintainability than LLM+function calling. Compared with the pure LLM, 20\% of participants require performance optimization (reducing latency/bandwidth consumption) and improved collaboration capabilities (enhancing serialization processing and intelligent scheduling). 10\% of senior engineers hold the view that MCP response latency is negligible compared to LLM latency, providing a different perspective on MCP performance evaluation. 90\% recognize improved output quality and 60\% recognize reduced monthly cost, driven by broken LLM knowledge boundaries, decreased invalid token consumption, simplified development, and lower human resource costs (Fig.~\ref{fig:mcp_core_dimension_changes}).

\begin{figure}[htbp]
    \centering
    \includegraphics[width=\linewidth]{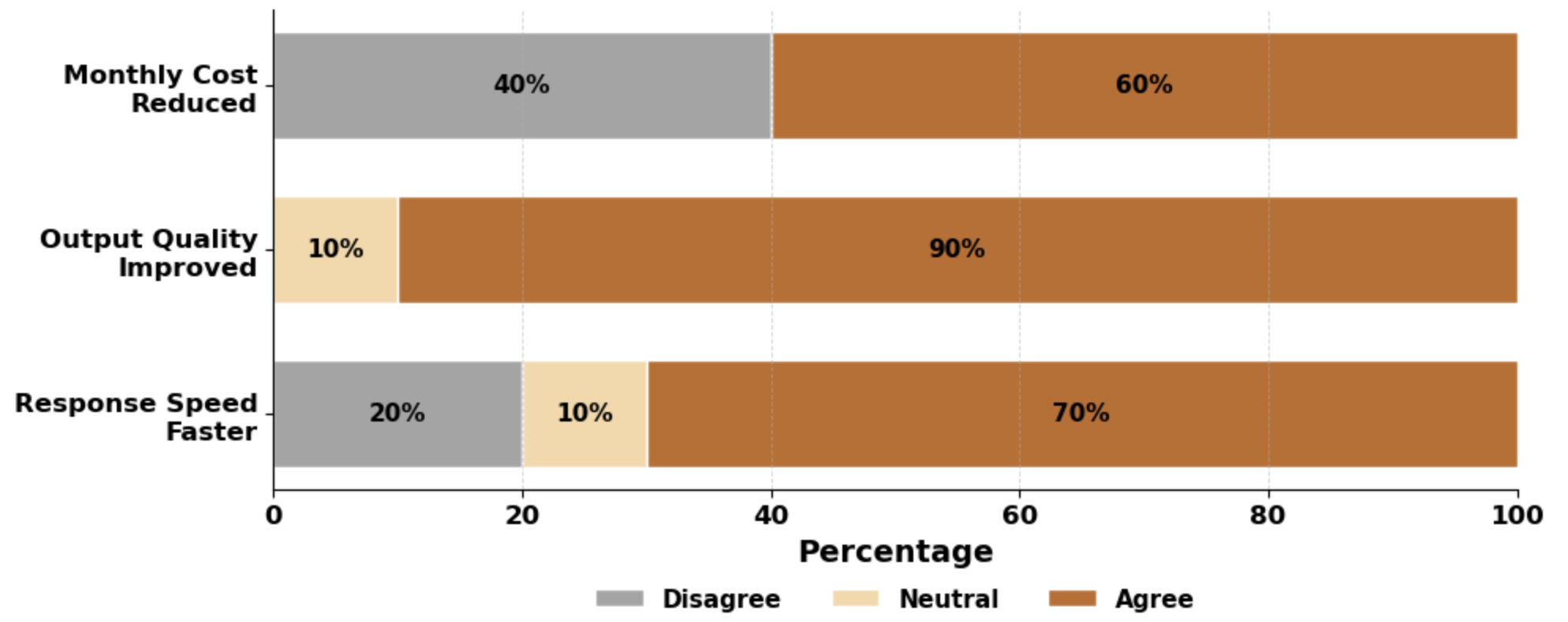}
    \caption{Views on MCP Performance Improvements}
    \label{fig:mcp_core_dimension_changes}
\end{figure}

In terms of technical combinations, architectures based on LLM+MCP and LLM+function calling exhibit distinct characteristics. For security risks, industry context plays a key role: FinTech participants reported almost no external interaction risks, whereas Internet participants faced a broader range of threats. As one practitioner noted: \textit{``Currently, in the fintech industry, the MCP service has bandwidth limitation problems. Essentially, it is a system protection mechanism rather than a problem at the model layer. Fault response is currently handled using degradation rules. The function of automatically switching models has not been implemented yet, and it takes about 10 minutes for manual switching''}(Participant 14). Among Internet practitioners, 80\% identified permission overstepping and prompt injection as MCP’s core security risks, with current defenses largely limited to fine-grained access control, audit logging, and API encryption, and lacking strong active protection mechanisms. All developers (100\%) agreed that MCP+LLM has slightly higher architectural complexity but lower development difficulty and better long-term maintainability than LLM+function calling. Compared with using LLMs alone, 20\% of participants emphasized the need for performance optimization (e.g., reducing latency and bandwidth consumption) and stronger collaboration capabilities (e.g., better serialization and intelligent scheduling). In contrast, 10\% of senior engineers considered MCP’s added latency negligible relative to LLM latency, offering a different perspective on performance evaluation. Overall, 90\% observed improved output quality and 60\% reported reduced monthly costs, attributed to breaking LLM knowledge boundaries, lowering invalid token consumption, simplifying development, and reducing human resource expenses (Fig.~\ref{fig:mcp_core_dimension_changes}). As participants noted: \textit{`` Development complexity: Using MCP will add a network interaction layer, leading to an increase in development complexity. If the MCP is maintained by others, the maintenance cost is relatively low; if it is developed by your own team, more communication and coordination are required''}(Participant 3).

\begin{center}
\vspace{-12pt}
    \resizebox{\linewidth}{!}{
\begin{tabular}{l!{\vrule width 1pt}p{0.9\columnwidth}}
    \makecell{{\LARGE \faLightbulbO}}  &\textbf{Finding 5.} %100\% of participants agree that the MCP+LLM architecture outperforms LLM+function calling in terms of long-term maintainability and development efficiency.
    Internet practitioners, in particular, highlight notable security risks in MCP-based systems, especially permission overstepping and prompt injection, and note that existing defenses remain largely passive. % 你参考下
\end{tabular}}
%\vspace{-8pt}
\end{center}

% ========== 插入结束 ==========

% %80% of participants reported that MCP–LLM architectures
% are more complex and require building an MCP server that
% integrates tool libraries and permission management. However,
% Fig. 5: Views on MCP Performance Improvements
% once MCP is in place, 90% believed that system scalability
% improves. All participants (100%) stated that LLMs using
% only function calling must be repeatedly adapted to differ-
% ent LLM platforms, which reduces development efficiency
% (Fig. 6). Over the long term, MCP–LLM combinations were
% viewed as more maintainable: participants noted that adding
% new tools typically only requires registration, which lowers
% maintenance effort and reduces rental costs for medium- and
% large-scale  . MCP-based architectures
% also have lower operating costs, help constrain improper oper-
% ations, standardize workflows, and support fault isolation rules,
% whereas LLMs with function calling alone require additional
% custom security logic. As one practitioner commented: “MCP
% combined with LLM requires more initial construction effort,
% but it is more flexible for adding new tools or expanding
% scenarios, making it more cost-effective in the long run”;
% another commented: “While LLMs with simple function calls
% have lower architectural complexity, cross-platform adaptation
% is cumbersome, causing maintenance costs to grow over time”. projects by about 30% 删掉了不合适量化，加参访编号

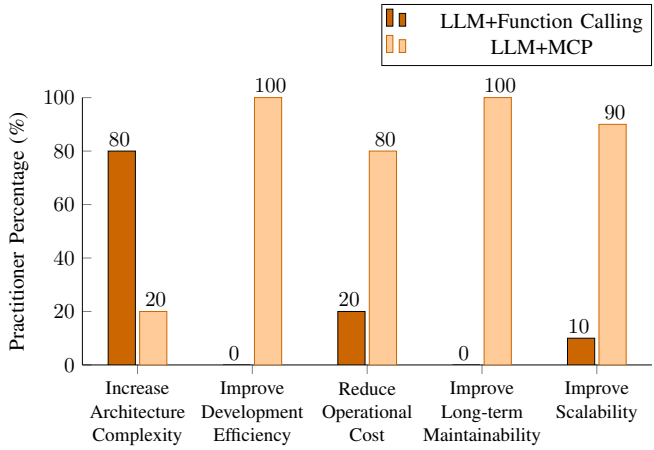
\begin{figure}[htbp]
\vspace{-10pt}
  \centering
  % 左移画布+微调缩放，拉近与左侧Y轴标题距离
  \begin{tikzpicture}[scale=0.8, transform shape, xshift=-1.2cm]
    \begin{axis}[
      width=11cm, height=6cm, % 宽度微调，适配左移后布局
      ybar,
      ymin=0, ymax=100,
      ylabel={Practitioner Percentage (\%)},
      ylabel style={font=\fontsize{10}{12}\selectfont, align=center}, 
      xlabel={},
      xlabel style={font=\fontsize{10}{12}\selectfont, align=center, yshift=-20pt},
      xtick={1,2,3,4,5},
      % 优化X轴标签分行，缩小字间距，彻底解决重叠
      xticklabels={
        Increase\\Architecture\\Complexity,
        Improve\\Development\\Efficiency,
        Reduce\\Operational\\Cost,
        Improve\\Long-term\\Maintainability,
        Improve\\Scalability
      },
      xticklabel style={
        font=\fontsize{9}{11}\selectfont, % 字号微降，适配布局
        align=center,
        yshift=0pt,
        inner xsep=0pt % 消除标签内边距，减少横向占用
      },
      bar width=0.45cm, % 柱子宽度微调，适配5个维度
      nodes near coords, 
      nodes near coords align=center,
      tick label style={font=\fontsize{10}{12}\selectfont, align=center},
      label style={font=\fontsize{10}{12}\selectfont, align=center},
      grid=minor,
      minor grid style={dashed, gray!30},
      axis lines*=left,
      enlarge x limits=0.12, % 微调X轴外侧间距，适配左移+5维度
      cycle list={orange!80!black, orange!40!white},
      legend pos=north east,
      legend style={
        font=\fontsize{10}{12}\selectfont,
        column sep=0.5cm,
        fill=white,
        draw=black,
        inner sep=2pt,
        at={(1.0, 1.35)},
        anchor=north east
      }
    ]
      % LLM+function calling 数据
      \addplot [fill=orange!80!black,
        every node near coord/.style={
          font=\fontsize{10}{12}\selectfont,
          yshift=0.2cm, 
          xshift=-0.3cm
        }]
        coordinates {(1,80) (2,0) (3,20) (4,0) (5,10)};
      
      % LLM+MCP 数据
      \addplot [fill=orange!40!white, draw=orange!80!black,
        every node near coord/.style={
          font=\fontsize{10}{12}\selectfont,
          yshift=0.2cm, 
          xshift=0.3cm
        }]
        coordinates {(1,20) (2,100) (3,80) (4,100) (5,90)};
      
      \legend{LLM+Function Calling, LLM+MCP}
    \end{axis}
  \end{tikzpicture}
  \caption{\fontsize{9}{13}\selectfont LLM+Function Calling \textit{\textbf{VS}} LLM+MCP}
  \label{fig:llm_mcp_func_comparison_rq3}
  \vspace{-14pt}
\end{figure}

80\% of participants reported that MCP–LLM architectures are more complex and require building an MCP server that integrates tool libraries and permission management. However, once MCP is in place, 90\% believed that system scalability improves. All participants (100\%) stated that LLMs using only function calling must be repeatedly adapted to different LLM platforms, which reduces development efficiency (Fig.~\ref{fig:llm_mcp_func_comparison_rq3}). Over the long term, MCP–LLM combinations were viewed as more maintainable: participants noted that adding new tools typically only requires registration, which lowers maintenance effort and reduces rental costs for medium- and large-scale projects. MCP-based architectures also have lower operating costs, help constrain improper operations, standardize workflows, and support fault isolation rules, whereas LLMs with function calling alone require additional custom security logic. As one practitioner commented: \textit{``MCP combined with LLM requires more initial construction effort, but it is more flexible for adding new tools or expanding scenarios, making it more cost-effective in the long run''}(Participant 2); another commented: \textit{``While LLMs with simple function calls have lower architectural complexity, cross-platform adaptation is cumbersome, causing maintenance costs to grow over time''}(Participant 5).

\begin{center}
\vspace{-12pt}
    \resizebox{\linewidth}{!}{
\begin{tabular}{l!{\vrule width 1pt}p{0.9\columnwidth}}
    \makecell{{\LARGE \faLightbulbO}}  &\textbf{Finding 6.} %Different industries (i.e., FinTech and the Internet) face distinct security risks associated with MCP, primarily due to the differences in their specific MCP application scenarios and compliance requirements.
    LLM+MCP architectures are more complex to build initially than LLMs with simple function calling, but practitioners consider them easier to maintain and more scalable in the long term. %你参考下 
\end{tabular}}
%\vspace{-8pt}
\end{center} 

The FinTech industry avoids external API connections, resulting in nearly zero risks of data leakage and external attacks. 70\% of participants require MCP tools to support local data processing. The Internet industry faces multiple risks: 80\% of participants are concerned about prompt injection and malicious tool invocation; 35\% worry about unauthorized data access; and sensitive word filtering struggles to adapt to diverse scenario requirements. Permission control serves as the core cross-industry protection measure; The FinTech industry adopts internal permission grading and API transmission encryption; the Internet industry implements fine-grained permission management through preset MCP access rules. Only 30\% of teams have achieved API-level fine-grained permission control, with most still relying on the simple ``administrator/regular user'' binary model. The Internet industry has added verification layers in the MCP data acquisition stage and implemented API transmission encryption, but only 20\% of participants reported deploying proactive defense mechanisms (e.g., prompt injection detection). Audit and monitoring mechanisms are inadequate, and the lack of real-time diagnostic tools hinders risk response speed. As one practitioner noted: \textit{``Current protection measures mainly rely on passive permission control, lacking effective detection methods against active malicious attacks''}(Participant 11).

\subsection{RQ4: What are participants' expectations for MCP, and what challenges occur in multi-model collaboration?}\label{AA}
This research question explores participants’ common and industry-specific expectations for MCP, while analyzing the causes of the ``stability superposition attenuation'' phenomenon in MCP multi-model collaboration scenarios.

Participants' expectations mainly focus on 3 aspects.
First, MCP should adopt a unified standard when integrated with different proxy frameworks, which can avoid the incompatibility issues currently existing in practical applications. 
Second, participants expect MCP to be more user-friendly and accessible through low-code or plugin-based approaches, thereby reducing the technical threshold for users without professional MCP development experience. 
Third, regarding fault repair and localization, participants hope that a standardized automated tool can be developed to detect and locate bugs in the MCP system, which will help improve the efficiency of fault handling and reduce the cost of operation and maintenance.
The ``stability superposition attenuation'' phenomenon refers to the decrease in system stability as the number of integrated tools increases, which stems from the combined effects of technical defects and resource constraints.

\begin{center}
\vspace{-12pt}
    \resizebox{\linewidth}{!}{
\begin{tabular}{l!{\vrule width 1pt}p{0.9\columnwidth}}
    \makecell{{\LARGE \faLightbulbO}}  &\textbf{Finding 7. }Participants expect MCP to be improved in three key aspects: unified standardization to resolve compatibility issues, low-code or plugin-based designs to lower usage barriers, and standardized automated tools to enhance fault detection, localization, and operational efficiency.
\end{tabular}}
%\vspace{-8pt}
\end{center} 

80\% of participants prioritize unified MCP interfaces, communication protocols, and access processes to reduce adapter development costs. Across both industries, there is demand for plug-and-play access and clear development guidelines. Security expectations vary by industry: FinTech focuses on data localization and internal permission management; the Internet industry needs unified prompt injection defense and multi-scenario sensitive word filtering. 70\% of participants call for the establishment of e MCP tool registration centers to avoid redundant development. Some participants commented: \textit{``Unified MCP standards would eliminate the need for adaptation across different frameworks, saving significant time''; ``We hope that multi-connection platform tools will be more convenient to use while enhancing security protection, for example, they can be deployed to the IDE like a plug-in''}(Participant 18).

According to statistics, 16 participants mentioned a lack of standardized diagnostic tools for MCP systems. Without dedicated tooling, fault localization and issue identification become extremely difficult in practice. Even simple failures require heavy manual effort to trace, and diagnostic logs are often incomplete or unstructured. As a result, both issue resolution and system recovery suffer from low efficiency and high uncertainty. Participants strongly expressed expectations for standardized, automated fault detection and diagnosis tools to help developers quickly locate bugs, analyze logs, and identify root causes during MCP operation and maintenance. Some participants commented: \textit{`` The automatic recovery mechanism of the MCP system is mainly stuck in the monitoring and detection link, which has not been perfected yet, resulting in low efficiency of fault location. Moreover, although the repair logic is clear, the lack of real-time diagnostic tools has dragged down the overall recovery efficiency''}(Participant 2).

\begin{center}
\vspace{-12pt}
    \resizebox{\linewidth}{!}{
\begin{tabular}{l!{\vrule width 1pt}p{0.9\columnwidth}}
    \makecell{{\LARGE \faLightbulbO}}  &\textbf{Finding 8.} Counterintuitively, although MCP is designed to integrate multiple LLMs for complex task execution, increasing the number of incorporated LLMs tends to reduce system stability. This is because the output instability of individual models can be amplified in multi-model collaboration, which in turn degrades overall system effectiveness.
\end{tabular}}
\vspace{-7pt}
\end{center} 

\textbf{Information redundancy and conflicts are direct triggers.} 65\% of participants reported that overlapping or contradictory results from multiple tools cause LLMs to enter self-verification loops. \textit{``It is believed that the MCP architecture will increase the complexity of operation and maintenance. One more service means one more potential failure point''}(Participant 20). Model adaptation differences amplify instability. Participants noted that \textit{``different models have different levels of support for the MCP protocol. Compared with the single-model scenario, the failure rate has increased by approximately 40\%''}(Participant 12). Context window limitations lead to key information loss, particularly prominent in early open-source small-window models. Scheduling logic flaws and performance degradation interact synergistically: unreasonable task allocation and keyword-based scheduling exacerbate issues; Participants noted: \textit{``The more MCP tools integrated, the more frequent result conflicts become, leading to decreased system stability''}(Participant 15); \textit{``Different models exhibit varying levels of MCP protocol support, often causing scheduling confusion during collaboration and impacting overall stability''}(Participant 14).
\section{Discussion}
\subsection{Research Implications: Core Values for the Scientific Research Community}\label{AA}

\textbf{(1) Strengthen the scenario-based and engineering design of MCP technology to make up for the shortcomings in industry adaptation research.} Existing MCP research does not combine the compliance requirements and business characteristics of FinTech, Internet, and other industries, resulting in a disconnect between technical results and actual deployment needs of enterprises. The scientific research community needs to conduct targeted research on industry-customized MCP technology, such as developing MCP adaptation modules for localized data processing for the FinTech industry, and designing lightweight cross-frame adaptation interfaces for the Internet industry to improve the industrial implementation of technological achievements.

\textbf{(2) Research on MCP distributed collaboration and status management needs to be systematically carried out to solve the core pain points of the ecosystem.} Existing research has not conducted an in-depth exploration of the collaboration mechanism under the MCP distributed architecture, leading to core problems such as ecological fragmentation, no native interruption mechanism, and confusion in keyword scheduling and identification in industry practice. The scientific research community should focus on researching MCP unified industry standards and interface specifications to break through the compatibility barriers of different Agent frameworks\cite{jin2024llms}.

\textbf{(3) Improve the MCP's ability to locate faults and improve the security defense system.} The scientific research community needs to build MCP fault location technology, build MCP-specific security threat data sets, and explore real-time detection and defense algorithms based on machine learning; at the same time, design security strategies based on industry characteristics, strengthen internal permission classification and data encryption capabilities for the FinTech industry, develop multi-scenario sensitive word filtering and malicious call identification mechanisms for the Internet industry, and design encrypted transmission and fine-grained permission control as MCP native core components.

\textbf{(4) Further explore the stability optimization of MCP multi-model collaboration and fill the technical research gaps.} Existing research has not paid attention to the ``stability superposition decay'' phenomenon in MCP multi-model collaboration, and the research on its formation mechanism, influencing factors, and optimization solutions is in a blank state\cite{jin2024llms}. The scientific research community needs to focus on research on MCP information redundancy filtering and dynamic resource allocation technology under multi-model collaboration, and analyze the mechanism of model uncertainty transmission and collaboration link amplification effect; at the same time, explore the integration optimization path of large models and MCP, use the small-sample learning ability of large models to analyze the causes of anomalies, solve problems such as response delays and result conflicts caused by multi-tool integration, and provide stability enhancement solutions for MCP systems with large-scale tool integration.

\subsection{Research Implications: Practical Guidance to Industry}\label{AA}

\textbf{(1) Build a standardized MCP implementation system to reduce technology use and operation, and maintenance costs.} Enterprises should establish a standardized MCP workflow covering ``tool registration-protocol adaptation-permission configuration-troubleshooting'' \cite{domanski2023digitalization} to clarify the responsibilities and division of labor of each role of Host, Client, Server and User; at the same time, build an  MCP knowledge governance platform, integrate tool manuals, common problem solutions, troubleshooting cases and other resources to promote cross-team knowledge reuse and avoid cost waste caused by repeated troubleshooting in new projects. In response to the problem of new fault points in MCP, enterprises need to establish a dedicated operation and maintenance monitoring system to shorten fault recovery time. At the same time, they should take advantage of MCP's standardized calling logic to speed up the training of new employees and lower the threshold for technology use.

\textbf{(2) Strengthen MCP full life cycle management and take into account standardization and customization needs.} Enterprises should establish a standardized MCP workflow covering ``tool registration-protocol adaptation-permission configuration-troubleshooting'', unify interface parameters, interaction protocols, and access processes to reduce cross-department and cross-project adapter development costs\cite{oberle2004developing}; At the same time, build an  MCP knowledge governance platform, integrate tool manuals, common problem solutions, troubleshooting cases, and other resources to promote cross-team knowledge reuse and avoid cost waste caused by repeated troubleshooting in new projects. In response to the problem of new fault points in MCP, enterprises need to establish a dedicated operation and maintenance monitoring system to shorten fault recovery time. At the same time, they should take advantage of MCP's standardized calling logic to speed up the training of new employees and lower the threshold for technology use.

\subsection{Threats to Validity}\label{AA}
One potential threat to the validity of our interviews is that some participants may fail to fully comprehend our questions. For instance, certain participants may be new to MCP and not understand of its various components. Consequently, they may not be able to clearly identify which MCP component they use in their work. To mitigate this threat, we provided an architectural flow chart of MCP to clarify the relevant questions. In addition, we allowed participants to respond with "I don’t know" when confronted with unfamiliar questions. These threats to validity are common in empirical studies\cite{wang2023practitioners,yang2017language} and can be mitigated through targeted measures.
Another potential threat to the validity of our research lies in our sample of practitioners; not all software engineers are covered. Specifically, our practitioners are limited to those who come from FinTech and the Internet and are proficient in English and Chinese, as well as those who have relevant open source projects on GitHub. Therefore, our findings may not fully represent the expectations of all software engineers. While we focus on studying the current usage of MCP and the problems encountered, there may be other aspects of MCP that we have not yet solved. We plan future research investigating these factors.

\section{Conclusions}

%This research adopted semi-structured interviews with multi-source data supplementation to investigate 20 participants from the FinTech and Internet industries. It for the first time completed a systematic empirical analysis of the practical application of MCP, exploring its application status, core value, key challenges and participants’ demands. This research fills the gap in existing studies regarding the actual situation of enterprise MCP deployment and provides important empirical support for the field of artificial intelligence system software engineering. Meanwhile, this research identifies the key challenges restricting the large-scale deployment of MCP. The conclusions of this paper provide practical guidance and directional insights for enterprises to deploy MCP applications and promote the technical iteration and standardized development of MCP. Future work will expand the sample to more industries such as manufacturing, healthcare and government to verify the universality of the conclusions, conduct long-term follow-up on the surveyed enterprises, and develop and pilot a low-code MCP prototype in collaboration with benchmark enterprises based on practitioners' expectations, so as to form a reusable standardized solution.

This study uses semi-structured interviews, supplemented with multiple data sources, to investigate 20 participants from the FinTech and Internet industries. It presents the first systematic empirical analysis of MCP's practical application, examining its deployment status, core value, key challenges, and practitioners’ needs. The study fills a gap in prior work on enterprise MCP deployment and provides empirical evidence for research at the intersection of artificial intelligence and software engineering. It also identifies critical obstacles to large-scale MCP adoption, offering practical guidance and strategic insights for enterprises planning MCP deployments, as well as for the iterative refinement and standardization of MCP itself. 
In the future, we plan to conduct follow-up interviews with a subset of the participating enterprises to track concrete changes in their MCP deployment practices over time, such as architectural adjustments, toolchain evolution, and refinements to security policies. 
%Future work will extend the sample to additional sectors such as manufacturing, healthcare, and government to assess the generalizability of these findings, conduct longitudinal follow-up with the participating enterprises, and co-develop and pilot a low-code MCP prototype with benchmark partners based on practitioner expectations, aiming to produce a reusable, standardized solution.

\bibliographystyle{unsrt}
\bibliography{ref}{}

\balance

\end{document}